# An Introduction to Digital Signature Schemes

Mehran Alidoost Nia[#1], Ali Sajedi[#2], Aryo Jamshidpey[#3]

[#1] *Computer Engineering Department, University of Guilan-Rasht, Iran*
m.alidoost@hotmail.com
[#2] *Software Engineering Department, Islamic Azad University- Lahijan Branch, Iran*
sajedi@iau-lahijan.ac.ir
[#3] *Computer Engineering Department, University of Guilan-Rasht, Iran*
aryo.jamshidpey@gmail.com

*Abstract*

Today, all types of digital signature schemes emphasis on secure and best verification methods. Different digital signature schemes are used in order for the websites, security organizations, banks and so on to verify user's validity. Digital signature schemes are categorized to several types such as proxy, on-time, batch and so on. In this paper, different types of schemes are compared based on security level, efficiency, difficulty of algorithm and so on. Results show that best scheme depends on security, complexity and other important parameters. We tried simply to define the schemes and review them in practice.

*Keywords*:
**Digital signature, Signature schemes, Security, Verification methods**

## 1. INTRODUCTION

Digital signatures are very important tools to implement secure and correct signs. Today, traditional physical signature is out-dated. Communications between partners of a company is significant issue that must be secure. Digital signature provides suitable background for sending secure messages using different schemes. Depending on different usages we must choose correct and appropriate option for signing our messages such as proxy-schemes. In this paper, we review and compare some of these implementation methods to optimize signing procedure.

As we mentioned above, usually digital signature schemes are categorized in 4 aspects [1]:

1. Schemes with Increased Efficiency
2. Schemes with Increased Security
3. Schemes with Anonymity Services
4. Schemes with Enhanced Signing and Verification Capabilities

Our research is about comparing these schemes in different scopes. For example, proxy signature is useful scheme for choosing alternative signers and delegation to a second person. This method improves group working and communication between company personals. This scheme is created again as "multi-proxy" signature scheme. [3] in this referenced paper stages of multi-proxy implementation is explained.

When a digital signature is generated, important thing is the security of signature scheme. Other parameter to evaluate digital signature scheme is difficulty of implementation. This parameter based on computer platform, used locations, security of scheme and so on.

It's trade-off. When you focus on the security, may miss speed and similarly when focus on Efficiency may implement more difficult than before.

## 2. LITERATURE REVIEW

Let us first define "Digital Signature". What is a digital signature and its applications?

*Definition1: Digital signature*
Digital Signature is a cryptographic primitive which is fundamental in authentication, authorization and non-repudiation. [2]

Digital signature proves its owner identity and he or she can't repudiate his or her sign. Digital Signature is implemented by public and private key algorithms and hash functions.

As we showed in Fig 1, when Original message is generated by user and sent for signing, Message is hashed and after performing private-key algorithm, digital signature is generated and is appended to message as "digital signature".

When user receives signed message, he or she can ensure from its validity. This procedure called *verification* that is performed by *public-key* algorithms.

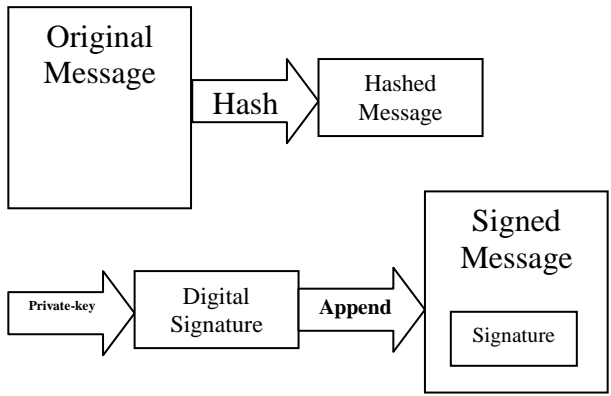

Fig 1 steps of generating Digital Signature

## 3. DIGITAL SIGNATURE SCHEMES

In this section we defined types of Digital Signature Schemes and procedures of their implementation. For beginning we must refer that in this paper we only introduced and compared these schemes and only we referenced to implementation algorithms.

*A. Batch Scheme*
This scheme is classified on Increased Efficiency [1] that provides synchronous signing with number of large scale computations. For example suppose that we need thousands of signing verification in the same time, in this case Batch Digital Signature Scheme is suitable.
Batch is proposed a procedure for signing Digital messages that are batched together at first. After performing different cryptographic algorithms, messages are signed and split in single signed messages that can be sent to requestors.
This scheme is used random numbers to generate and verify validity of signers. This specific of Batch Scheme is caused that attackers can't to access these numbers. For reconstructing verification functions, attacker is needed to know these random numbers but it seems infeasible in practice. For introduction with its algorithms read referenced paper. [1]

*Definition2: RSA algorithms*
Suppose that given a positive integer *n* and two odd prime numbers *p* and *q*.
Positive *e* selects that gcd (e, (p−1) (q−1)) = 1 and an integer c. we must find an integer m such that $m^e = c \pmod{n}$.[2]

RSA algorithm is used in many types of schemes and is a significant basic approach to implement Digital Signature Schemes.

*B. Forward-Secure Scheme*
It is categorized in *Schemes with Increased Security*. Its security level is very high because maintains validity of the key, after key is comprised [1]. As we referred its high level security is caused to large of usages. Idea behind Forward Secure Scheme is T (total time of verifying public key). Splits time T into equal periods that any periods have special different Secret Key.
Public key is remained constant while next Secret Key is generated according to previous Secret Key and *Key Update Algorithm*. In special way, generates two random & prime numbers *P1* & *P2* and uses them to generate and verify steps of signature generation and verification.

**Definition 3: Update Algorithm**
It's a simple algorithm to calculate signatures in periods of time.
1. If j=T then return to generation protocol
2. Regenerates $e_{j+1}$, ... , $e_T$ and starting with constant P
3. Computes $S_{J+1}$

That *j* is the current time; *S* is the generated Signature in this period.

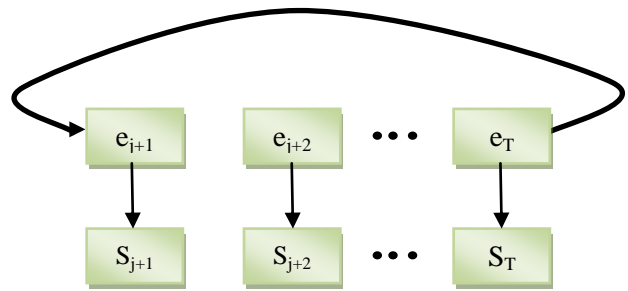

*Fig 2 Forward-Secure implementation steps*

*C. Blind Scheme*
This scheme is classified on *Anonymity Services* [1] that in this approach sender must get sing from signer but signer can't know sender identify and only sign this message.
Suppose that person x sends a message for signing to the person y and y must signs this and sends it again to sender without knowing sender's identity. This implementation includes of three steps that should be executed sequent.

First stage called *Blinding*: consists of changing message *m* into the form of $f(m)$. Function $f$ is the blinding function. This action changes message's identity.

Second stage called Signing: in this stage person y sign the message and send it back to sender. But message changes to $S(f(m))$ that can't reveal without sender's key.

Last stage called Unblinding: function g is performed to the message and sign will verify.
$g(S(f(m))) = S(m)$

Now sender can use and see the authentication of signer. According to the Fig 3.

### D. Proxy Scheme

This scheme is classified on *Enhanced Signing and Verification Capabilities*. [1] Gives privilege of signing to another person that signer trusts him. For example president of company can't stay in office and can't signs messages for end of weeks. He must delegates privilege of signing to his secretary or vice president of company. Proxy scheme is provided this work's capability. In this idea permission is granted to a person but original sign is not revealed for him.

This scheme is required a complement scheme for implementation and its security and other parameters depends on complement algorithm.

## 4. COMPARING THE SCHEMES

The mentioned schemes are compared regarding in different aspects. Significant parameter for computing and using Schemes is the security of algorithm implementation. We focus on security parameters at first.

### A. Security

*Batch* Scheme's security is very strong but not *sound*.

### Definition 4: Sound

Called in approaches that valid signatures may not identify as invalid and invalid signatures also may not identify as valid signatures. [1]

Batch schemes are very strong but in several conditions makes mistake to identify signature verification.

*Forward-Secure* schemes are focused more on security parameters and its users claim that anyone can't to break it. Because of its different generation keys in several periods of time, if attackers know one step's key, he can't access to original signature generator. Random and prime numbers that used in its algorithm warranty the total security.

Blind scheme's security depends on two parameters. Blindness & unforgeability [1]. Blindness means that signer can't identify sender's identity. It can be useful and can be harmful. It's can be harmful because of attackers. Attacker's identity can't be reveal for signers and signer only signs the message.

Unforgeability is a good parameter for enhance security. After sender gets the blind-signed message he can't access to original signature protocol. Because the pair of (message-sign) only can be generated by signer. Suppose that attacker access to signed messages but he can't calculates original until has Public-key that used in calculating formula of Digital Signature generation.

Proxy Scheme security depends on complement algorithm that is used for implementation. Generally, we used *ElGamal* signature scheme [1] to implement proxy schemes. It seems infeasible an attacker can break proxy. But in practice it's proved that an invalid proxy signer could cheat protocols and generate proxy signed message.

A solution for this problem is *Key Delegation Protocol*. By adding this protocol to proxy, we prevent probability attacks and invalid messages.

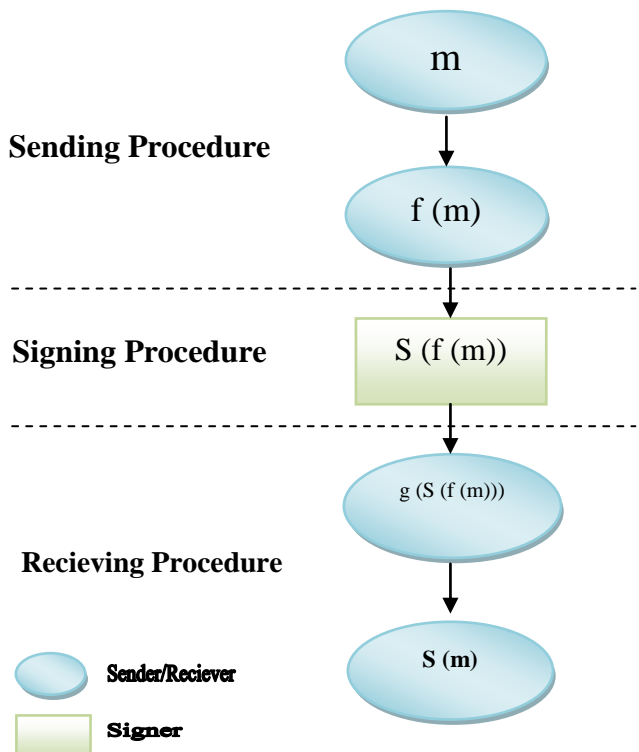

*Fig 3 Blind scheme implementation steps*

*TABLE 1 comparing Digital signature schemes*

| Aspects / Scheme | Security | verification | difficulty | Efficiency |
|---|---|---|---|---|
| **Batch** | Middle | middle | Low | Low (individuals are more efficient in some applications) |
| **Forward secure** | Very strong | Very good | middle | efficient (time consuming) |
| **Blind** | strong | good | middle | Very efficient |
| **Proxy** | Strong (based on complement algorithms) | good (based on Key Delegation protocol) | high | efficient |

### B. Verification
Verification shows how much sings of these schemes are verified correctly and gives us the validity and invalidity of signatures in practice. When a message is signed and is sent to desired location, receiver verify message with his or him public-key. If decryption procedure is done correctly, verification is acceptable. If not, identity of signer is denied.

Procedure of verification must certify signer's identity. In some cases, May a valid signature is recognized as invalid and an invalid signature is recognized as valid. This is a bad feature for verification that its result always is not true.

### C. Difficulty
Difficulty is the parameter that indicates easiness of implementation for programmers. This parameter is real and gives the programmers an option to choose schemes according to their skills of programming. In some platforms, this is important to decrease number of calculations depends on hardware specifications. As a programmer, we must choose options. For example if computer hardware is designed professionally, the programmer is free to choose level of difficulty. As mentioned in *TABLE 1*, difficulty is categorized in three levels. Easiest level for programming is *low* and then *middle* & *high*.

### D. Efficiency
Efficiency parameter defines which scheme is more effective and optimal for its applications. This is an approximation parameter that based on several conditions such as programmer's skill to implement mentioned algorithms, security of network and system designs, percentages of verification signatures, relations between choosing appropriate scheme and its applications and so on.

As referred above, efficiency is calculated approximately and it's not enough for judgement among these schemes. This only provides an overview to select appropriate schemes according to our applications.

In this section we compared mentioned schemes that are showed in *TABLE 1*.

## 5. CONCLUSION & FUTURE WORKS

Today, everyone try to improve and combine good parameters of all schemes to achieve a new and secure scheme. For example, Zhang, et all have worked on security of proxy by enhancing security level as mentioned. [4] Another example that is offered by Zhenhua, et all is used multi-proxy signature scheme with revocation. [3] They proved that multi-proxy scheme is designed by them, is secure against a *type 2 & 3* adversary. As indicated they worked on security in high level.

Security parameter shows the Security level of the scheme and helps us to select appropriate scheme according to applications. This parameter is more significant than other mentioned parameters in *TABLE 1*. Scheme's capability depends on security of signing algorithm.

However, this is a trade-off. When the emphasis on security, we must program signature scheme difficultly and if we rely on low difficulty of implementation then we miss the security.

Approximately all this parameters are correlated. However, the following enhancement can be accomplished to introduce works that we can involve in future. A lot of schemes are based on security and enhance this parameter with focus on block attacker's invasions. Combining two or more of schemes on different cases can be useful and its result is increased security or optimizes algorithms that must generate Digital Signatures. Adding random oracles model that offered by Jianhong et al , changing proxy schemes to multi-proxy schemes [3], Fast certificated-based algorithms by optimizing signature scheme's implementation algorithms that worked by Levente et al [6], identify-based Signcryption schemes that focused on verification methods as mentioned [5] and similar works that perform new generation methods can be optimized and enhances security level in different usages are some aspects for future works.